\documentclass[conference]{IEEEtran}
\IEEEoverridecommandlockouts
\usepackage{amsmath,amssymb,amsfonts}
\usepackage{algorithmic}
\usepackage{graphicx}
\usepackage{textcomp}
\usepackage{multirow}
\usepackage{multicol}
\usepackage{wrapfig}
\usepackage{etoolbox}
\usepackage{biblatex}
\usepackage{caption}
\usepackage{subcaption}
\usepackage{xcolor}
\def\BibTeX{{\rm B\kern-.05em{\sc i\kern-.025em b}\kern-.08em
    T\kern-.1667em\lower.7ex\hbox{E}\kern-.125emX}}
\begin{document}

\title{SoK: Public Blockchain Sharding\\
\thanks{This work was supported in part by the US National Science Foundation under grants 2247560, 2154929, 1916902, and 2331936, by the Office of Naval Research under grant N00014-19-1-2621, and by the Virginia Commonwealth Cyber Initiative (CCI).}
}


\author{\IEEEauthorblockN{Md Mohaimin Al Barat,
Shaoyu Li,
Changlai Du, 
Y. Thomas Hou, and
Wenjing Lou}
\IEEEauthorblockA{Virginia Tech, USA}
\IEEEauthorblockA{\{barat, shaoyuli, cdu, hou, wjlou\}@vt.edu}
}

\IEEEoverridecommandlockouts

\IEEEpubid{\makebox[\columnwidth]{979-8-3503-1674-2/24/\$31.00~\copyright2024 IEEE \hfill} \hspace{\columnsep}\makebox[\columnwidth]{ }}

\maketitle

\IEEEpubidadjcol

\begin{abstract}
Blockchain's decentralization, transparency, and tamper-resistance properties have facilitated the system's use in various application fields. However, the low throughput and high confirmation latency hinder the widespread adoption of Blockchain. Many solutions have been proposed to address these issues, including first-layer solutions (or on-chain solutions) and second-layer solutions (or off-chain solutions). Among the proposed solutions, the blockchain sharding system is the most scalable one, where the nodes in the network are divided into several groups. The nodes in different shards work in parallel to validate the transactions and add them to the blocks, and in such a way, the throughput increases significantly. However, previous works have not adequately summarized the latest achievements in blockchain sharding, nor have they fully showcased its state-of-the-art. Our study provides a systemization of knowledge of public blockchain sharding, including the core components of sharding systems, challenges, limitations, and mechanisms of the latest sharding protocols. We also compare their performance and discuss current constraints and future research directions.
\end{abstract}


\section{Introduction}
Since its inception in 2008, blockchain technology has significantly transformed the digital world. Initially introduced as the foundational data structure for the first successfully deployed cryptocurrency, Bitcoin \cite{bitcoin}, blockchain technology has since been extensively adopted across various sectors, including cryptocurrencies \cite{ethereum}, medical area \cite{medical,medical2}, Internet-of-Things \cite{iot,iot2}, government sectors \cite{govt,govt2}, artificial intelligence \cite{artin,artin2}, decentralized finance (DeFi) \cite{defi}, decentralized applications (dApps) \cite{dapps}, and others \cite{others}. Each of these sectors gains advantages from blockchain's transparent, decentralized, and immutable qualities, along with its fully distributed peer-to-peer architecture, which is essential for recording digital tokens like transactions. The blockchain functions as a database that records every transaction within the blockchain network, with each participating node maintaining a replicated copy. It establishes a decentralized ledger, eliminating the need for a central authority to establish trust or validate transactions. Trust is not assumed among participating nodes, making blockchain a tool for enabling secure computation among mutually distrustful participants. Additionally, blockchain is renowned for providing a reliable and unchangeable record-keeping service. The transactions are validated by the \textit{miners} or \textit{validators} of the blockchain network, who then add these transactions into new blocks. The block data structure in the blockchain includes the hash of the previous block in the subsequently generated blocks. This use of a hash chain ensures that altering data in one block would render later blocks invalid \cite{xiao2020survey}. Since the system is \textit{distributed} and \textit{decentralized}, the nodes need to reach an agreement regarding the state of the ledger or the validity of the transactions. Several consensus protocols have been introduced to achieve this goal, including \textit{proof-of-work} \cite{bitcoin}, \textit{proof-of-stake} \cite{peercoin}, \textit{byzantine fault tolerence} \cite{bftcon}, etc.

\textit{Scalability} is a critical factor for the widespread adoption of blockchains that aim to provide high-quality services to the general public across expansive networks with unlimited growth potential. A scalable blockchain provides a better user experience with better performance (faster transaction times and lower costs). Moreover, as the adoption of blockchain increases, the number of transactions also rises. Scalability ensures that the blockchain can handle this increased volume without significant delays or increased costs.

However, low scalability is one of the major challenges that blockchain faces. To maintain \textit{decentralization}, the nodes in the network need to reach a consensus on new transactions or blocks. This process, while securing the network and ensuring trustlessness, is inherently slower than centralized systems. Besides, traditional blockchains like Bitcoin have a limit on block size and the frequency at which blocks are added to the chain \cite{bitcoin}. These limitations, initially designed to maintain network security, can lead to lower throughput and scalability as the number of transactions increases.
The low throughput of current blockchain-based cryptocurrencies (7 transactions per second (TPS) for Bitcoin \cite{tps} and 15 TPS for Ethereum \cite{tps_ether}, compared to Visa average 1776 TPS and PayPal average 700 TPS \cite{visaandpaypal}) presents a substantial bottleneck that limits their scalability and wider practical application.

Several methods have been proposed to enhance the scalability of blockchain systems, which can be categorized into two groups: first-layer solutions and off-chain solutions. First-layer solutions refer to those directly applied to the main blockchain or its consensus protocol \cite{sok_layer2}. Examples include increasing block size to accommodate more transactions per block \cite{bitcoin_cash}, implementing directed acyclic graphs (DAG) \cite{dag}, and exploring alternatives to Proof of Work (PoW) such as EOS and Stellar \cite{eos,stellar}. Additionally, to cope with the impact of a faster block generation time, Ethereum introduced the Greedy Heaviest Observed Subtree (GHOST) protocol, organizing blocks in a tree to enhance blockchain performance \cite{ghost}. Subsequently, the GHOST protocol was expanded to leverage a directed acyclic graph (DAG). Moving to the second category, off-chain solutions handle user requests outside the main chain \cite{sok_layer2}. These solutions mitigate latency by processing transactions off-chain and then settling them on the blockchain. Off-chain protocols like sidechains \cite{plasma}, rollups \cite{zkrollups} or payment channels \cite{raiden,lightning} manage transactions, minimizing interactions between nodes and the blockchain. However, these methods are not flawless. Off-chain approaches are more susceptible to forks \cite{layer2}, and transactions in a DAG setup do not adhere to a traditional chain structure.

In addition to the aforementioned strategies, blockchain sharding is an effective method for enhancing blockchain performance. The sharding system divides the blockchain network into \textit{shards} or \textit{zones} where nodes are allowed to store and process transactions of a single shard or multiple shards \cite{survey_sharding}, not from the whole blockchain network. The advantage of sharding is that when the whole blockchain network is divided into multiple shards, the nodes in the shards can work in parallel to validate the transactions and add them to the blocks. By leveraging the sharding technique, the performance of the entire blockchain can increase linearly \cite{elastico} with the increasing number of nodes, which enables partial transaction processing and storage on a single node. Among the aforementioned scalability solutions, sharding methodologies stand out as particularly promising, as they effectively address both performance (throughput and confirmation latency) and scalability challenges \cite{scaling}. Several sharding protocols have been proposed to explore horizontal scaling, which supports higher throughput and security mechanisms but also has some limitations. 
Compared to other methods, blockchain sharding has distinct advantages. For example, the first-layer solutions need to modify the protocols of different blockchains like increasing the block size, which will bring the network burden to the systems and some further negative impacts, while the blockchain sharding technology is able to reduce network burden and allows messages to reach their destination in a timely manner. 
The off-chain methods require executing transactions outside the blockchain, preventing nodes from directly interacting with blockchains to send transactions. This contradicts the fundamental design principle of cryptocurrencies. In contrast, blockchain sharding does not necessitate a reduction in interactions.

From what we understand, current research, as referenced in studies like \cite{sok, survey_sharding}, has not fully explored the important aspects of blockchain sharding. There is a noticeable gap in research that thoroughly investigates the main elements of this technology. Additionally, there is a lack of detailed comparison between the newer and more advanced sharding protocols. It is also important to point out that these studies have not deeply analyzed the drawbacks and performance standards of these protocols. This suggests that there is a significant need for more extensive research that covers these areas, providing a clearer picture of where blockchain sharding stands today and where it might be heading.

The goal of this paper is to offer a well-organized and detailed look at blockchain sharding. We will delve into the key parts and explain how they work. We are also comparing newer and more advanced sharding protocols in blockchain technology. This includes looking closely at how they work, their performance and limitations, and their future aspects. We have organized our paper to focus on the most important parts of these sharding protocols. In each section, we will lay out the problems these protocols are trying to solve, the objectives, and how these protocols tackle these issues. This approach should give a thorough understanding of blockchain sharding, both in its current state and in the possibilities it holds for the future. 

The organization of the paper is as follows. Section \textrm{II} provides an overview of blockchain sharding and its components. Section \textrm{III} discusses the first key component, which is identity establishment and committee formation. Section \textrm{IV} discusses consensus mechanisms. Section \textrm{V} discusses cross-shard transactions. Section \textrm{VI} discusses epoch randomness and reconfiguration of the committees. Section \textrm{VII} compares the performance of the sharding protocols. Section \textrm{VIII} provides the discussion and future research trends. Section \textrm{IX} concludes the paper. 

\section{Sharding Overview and Challenges}
Sharding is a concept borrowed from database systems. In database management, sharding is basically a division process where a large dataset or database is divided into smaller pieces to make it more manageable, and these smaller parts are called data shards \cite{db_horizontal_sharding}. The term \textit{shard} refers to a small portion of the whole dataset, and {\textit{horizontal partitioning} is referred to as \textit{sharding}. Sharding aims to make a large database more manageable by dividing it into smaller, faster parts. In terms of database, blockchain can be referred to as a decentralized database \cite{byteball} where all the full nodes store the data locally. Decentralized databases focus on distributing control and storage across multiple nodes for enhanced security and fault tolerance, often in a blockchain context. In a centralized database, sharding means splitting the data into multiple chunks to store it across one or multiple servers. In the decentralized database, or we can say, in blockchain, sharding involves dividing the network into smaller committees to process and store the blockchain data (Figure \ref{sharding_overview}). The number of committees increases in proportion to the network's overall computational capability. The goal of sharding in blockchain is to reduce the redundancy and overhead of communication, storage, and computation. Sharding is now considered one of the best ways to build a scale-out system to manage parallel processing, storage, and computation. 

\subsection{Sharding Overview}
Sharding protocols proceed in epochs, where the key idea is to parallelize the resources and divide the whole network into smaller groups or committees, each processing a disjoint set of transactions. Each of the nodes in the network needs to establish an identity to join the committees, and there are several mechanisms to do it, like proof-of-work \cite{bitcoin}, proof-of-stake \cite{peercoin}, and proof-of-personhood \cite{personhood}. The nodes are assigned to committees, and the committees reach an agreement within their zones to append blocks and transactions to the chain. The committees are reshuffled after each epoch to prevent adversaries (such as attacks on any specific committee) and biases (so that the adversaries can not gain any advantage in being assigned to the committees). To maintain decentralization while maximizing throughput, the number of committees in the network increases in proportion to the number of nodes. This ensures that transactions and blocks can be processed in parallel, resulting in increased throughput.

Many blockchain sharding protocols have been proposed in the literature. This paper focuses on sharding in public blockchains, where scalability poses a greater challenge than in private blockchains. In brief, \textit{Elastico} \cite{elastico} is the first sharding protocol that can increase throughput while ensuring decentralization and security of the network. Later, \textit{OmniLedger} \cite{omni} introduced \textit{state blocks} to mitigate the need to store the whole ledger for the nodes and used \textit{Atomix} to handle cross-shard transactions, the transactions that involve multiple shards for inputs and outputs rather than a single shard. \textit{RapidChain} \cite{rapid} was the first sharding protocol to maintain a disjoint ledger for each shard. This protocol also introduced \textit{decentralized bootstapping} in an untrusting setup. \textit{Monoxide} \cite{mono} introduces eventual atomicity for cross-shard transactions and Chu-ko-nu mining to ensure robustness against adversaries. The TEE-based sharding protocol \cite{towards} supports workloads beyond cryptocurrency. \textit{Pyramid} \cite{pyramid} is a layered sharding protocol that reaches intra-shard consensus and handles cross-shard transactions with two types of shards. \textit{Repchain}  \cite{repchain} sharding protocol uses \textit{reputaion scores} with a double-chain architecture. \textit{BrokerChain} \cite{brokerchain} introduces \textit{account segmentation} and \textit{state partitioning} to reduce the number of cross-shard transactions, and uses \textit{broker accounts} to handle cross-shard transactions. 

\subsection{Notations}

The notations used in the following content are presented in TABLE \ref{notations}. For the rest of the paper, \textit{shard}, \textit{zone} or \textit{committee} are interchangeable unless mentioned otherwise. 

\begin{table}[h]
\centering

{
\begin{tabular}{||c|c||}
\hline
Notation & Definition                     \\ \hline \hline
$n$      & Number of nodes in the network \\ \hline
$k$      & Number of shards               \\ \hline
$m$      & Number of nodes in each shard  \\ \hline
$H(x)$   & Hash of $x$                    \\ \hline
$h_i$    & Header of block $i$            \\ \hline
$Tx$     &         Transaction            \\ \hline
$T$      & Timestamp                      \\ \hline
$e$     & Epoch Number                    \\ \hline
$\epsilon$ & Epoch randomness           \\  \hline
\end{tabular}
}
\caption{Notations}
\label{notations}
\end{table}

\subsection{Problem Definition}
We follow the definition by \textit{Elastico} \cite{elastico}. Assume that there are $n$ nodes in the blockchain network, and each node has the same computing power. As in many theoretical models and simulations, this equal computing power is a common assumption made to simplify the analysis and focus on the core aspects of a problem. A Byzantine adversary has control over a portion $f$ of these nodes. A transaction $i$ in block $j$ is represented by an integer $x_i^j \in \mathbb{Z}_N$ in the ring of integers modulo $N$, denoted by $\mathbb{Z}_N$. To ascertain whether each transaction is genuine, each node has access to a constraint function, $\mathcal{C} : \mathbb{Z}_N \rightarrow \{0, 1\}$, that has been externally specified. The sharding protocol looks for a protocol called $\Gamma$ that runs across nodes and produces a set called $X$ that has $k$ different \textit{shards} or subsets, $X_i = \{x_i^j\}(1 \leq j \leq |X_i|)$, such that the following conditions hold:
\begin{itemize}
    \item Agreement: For a given security parameter $\lambda$, honest nodes concur on $X$ with a probability of at least $1-2^\lambda$.
    \item Validity: The given constraint function $\mathcal{C}$ is satisfied by the agreed-upon shard $X$, i.e. $\forall i \in \{1...k\}$ and $\forall x_i^j \in X_i$, $\mathcal{C}(x_i^j) = 1$.
    \item Scalability: The size of the network has an essentially linear effect on the value of $k$.
    \item Efficiency: The amount of computation and bandwidth required for each node to participate in the sharding protocol does not increase as the number of nodes or shards grows.
\end{itemize}
Sharding's goal is to divide the network into various committees, each of which handles different transactions (represented by a shard). Due to the almost linear increase in the number of shards as the network grows, smaller committees can execute consensus procedures more efficiently. Regardless of the number of nodes and shards, the amount of computing and bandwidth used per node remains constant. A distributed ledger of transactions is created in the blockchain when the network decides on a set of transactions and forms a hash chain with other previously agreed-upon sets from earlier iterations of the sharding protocol.

\begin{figure}
    \centering
    \begin{subfigure}[b]{0.45\textwidth}
        \includegraphics[width=\textwidth]{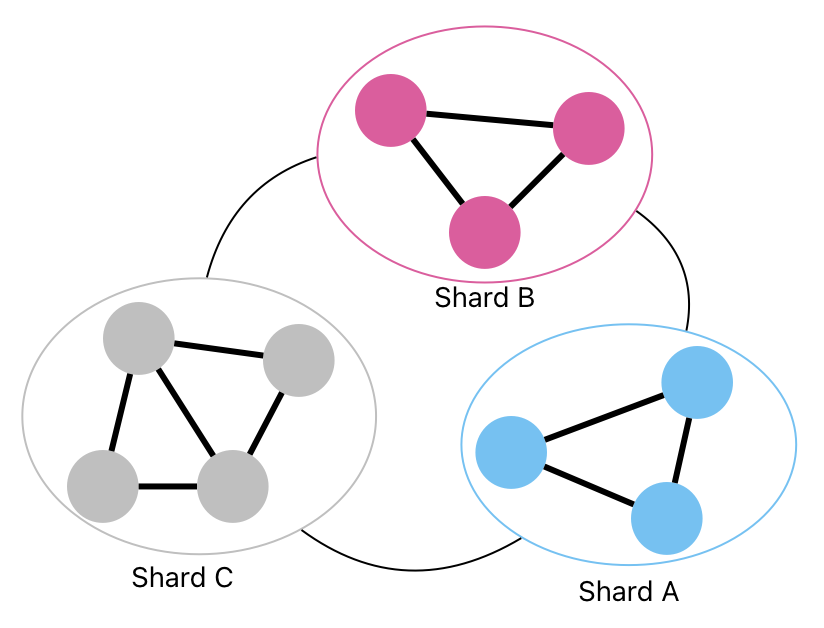}
        \caption{Sharded Network}
        \label{overview_a}
    \end{subfigure}
    \hfill
    \begin{subfigure}[t]{0.45\textwidth}
        \includegraphics[width=\textwidth]{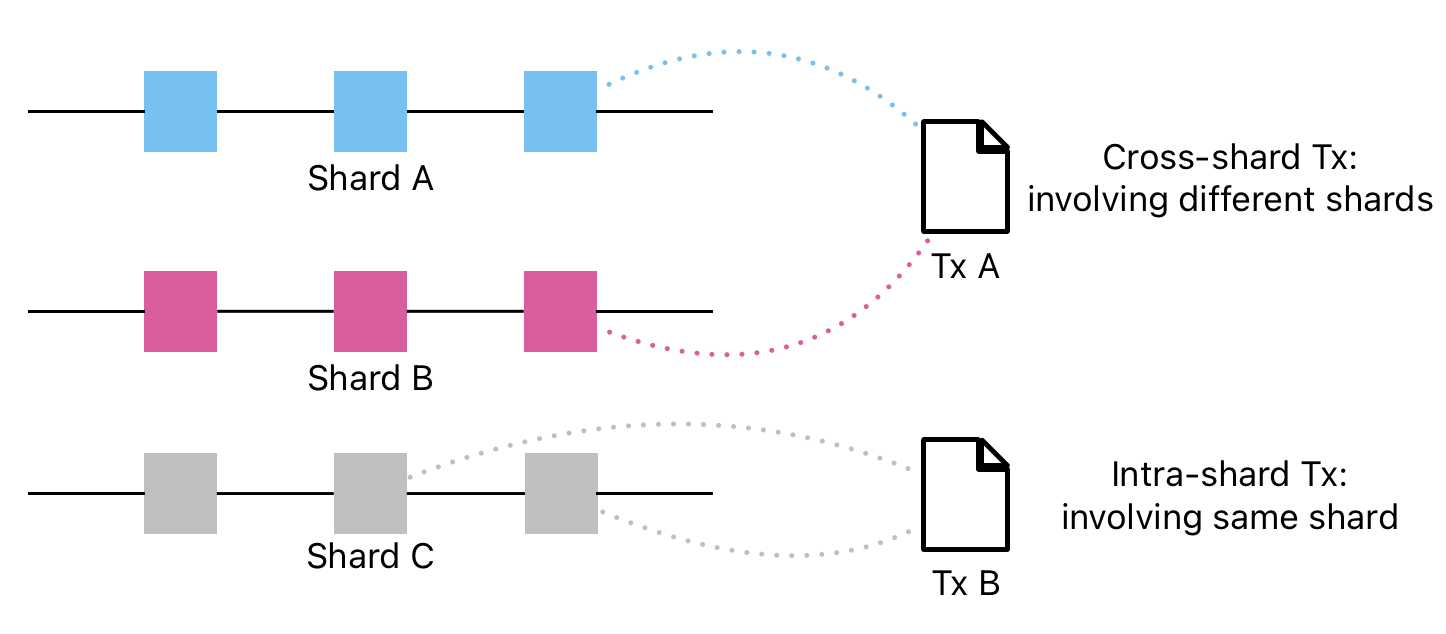}
        \caption{Sharded Ledger and Corresponding Transactions}
        \label{overview_b}
    \end{subfigure}
    \caption{Figure (a) shows the divided network into multiple shards. Figure (b) shows how the shards maintain disjoint ledgers, the intra-shard transaction involving only one shard, and cross-shard transactions involving multiple shards}
    \label{sharding_overview}
    
\end{figure}

\subsection{System Model}
The system model of a sharding system depends on the sharding mechanism of that protocol. In certain protocols, each shard carries identical responsibilities and follows the same set of procedures. In other protocols, shards are divided into different classes, each with unique (or some common) responsibilities. Based on these, we can effectively categorize sharding system models into two distinct groups: the \textit{homogeneous} model, where all shards operate in the same manner, and the \textit{heterogeneous} model, where shards have varied roles and functions. This classification not only simplifies our understanding of sharding systems but also highlights the strategic choices made in their design.
\subsubsection{Homogeneous Model} The \textit{validators} of the model handle the transactions and ensure the uniformity of the ledger. Suppose that the total number of \textit{validators} or \textit{nodes} is represented by $n$. Each validator or node $i$ has its own public-private key pair $(pk_i,sk_i)$. The network is divided into $k$ shards (where $k<n$), and each shard has $m=n/k$ nodes. The system works in a fixed time interval, which is called \textit{epoch} $e$. Depending on the system, each epoch can be a day or a few days long. Within each epoch, there are multiple \textit{rounds} $r$, and within each round, the nodes validate the transactions from the transaction pool and add valid blocks to the ledger, including the transactions. To participate in the epochs, the nodes can follow any Sybil-resistant mechanism (i.e., PoW) to establish their identity. 
\subsubsection{Heterogeneous Model} The shards in this system are divided into two categories. The \textit{i-shards} in Pyramid handle the intra-shard transactions. Each \textit{i-shard} can work independently to verify the transactions within its shard and add them to the ledger. \textit{b-shards} work as the cross-shard transaction handler, the transactions that involve multiple shards as input or output. The shards in BrokerChain are also divided into two categories but with different purposes. The mining shards or \textit{M-shards} in BrokerChain work as the transaction block generator and achieve intra-shard consensus. A partition shard or \textit{P-shard} handles the account state partitioning during each epoch. 

\subsection{Network Model}
The sharding network operates under some fundamental assumptions. Firstly, the network graph formed by honest validators exhibits strong connectivity. Secondly, the communication channels connecting these honest validators operate synchronously. Consequently, if an honest validator broadcasts a message, all other honest validators will receive the message within a predictable maximum delay $\Delta$ (optional to preserve message order). Additionally, each communication sent across the network undergoes authentication by utilizing the sender's private key, ensuring message integrity and sender verification. This synchronous communication is required only for the intra-consensus mechanism. For the other parts, partially synchronous channels are used to achieve responsiveness. 

\subsection{Threat Model}
In the sharding network, nodes are categorized into two types. The first type, \textit{honest nodes}, consistently exhibits cooperative behavior, adhering to protocols and collaborating with other honest nodes to achieve consensus. The second type is \textit{malicious} or \textit{corrupt} nodes. These nodes can disrupt the network's functionality by deliberately delaying communications, transmitting invalid requests, or attempting to manipulate transactions or blocks. The proportion of malicious nodes in the network at any given moment is represented by \(f \). This parameter `f' typically represents the maximum percentage of faulty or malicious nodes that the system can tolerate while still functioning correctly. In the case of the Monoxide protocol, this fraction is \( f = \frac{1}{2} \), while for RapidChain, it stands at \( f = \frac{1}{3} \). For other protocols, this fraction is typically \( f = \frac{1}{4} \). Another assumption is that the Byzantine adversary is slowly adaptive. So, the number of malicious and honest nodes only changes between each epoch or before the start of the protocol but can not be changed within each epoch. 

\begin{figure*}
    \centering
    \includegraphics[width=0.8\textwidth]{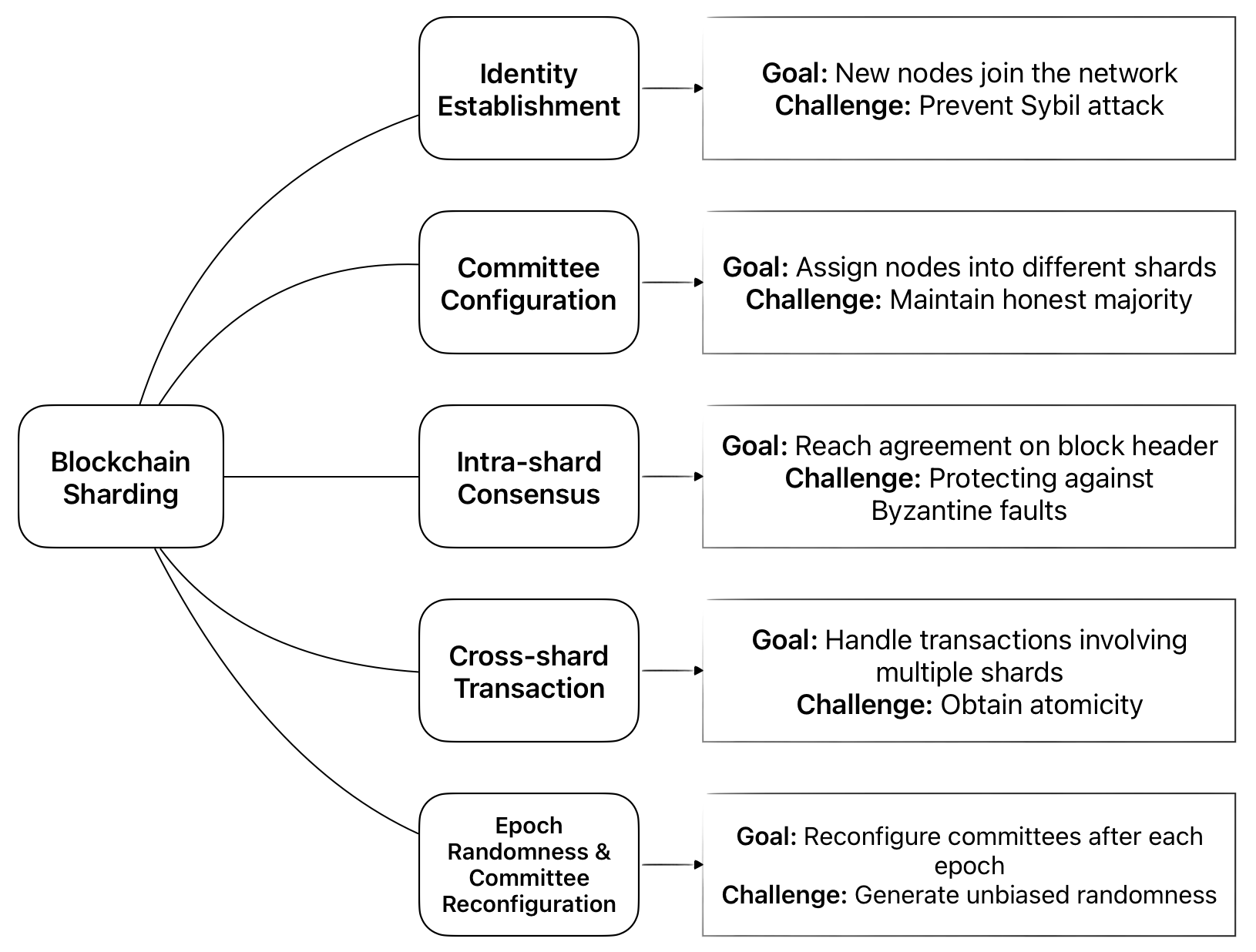}
    \caption{Key components of sharding and their goals}
    \label{fig:components}
\end{figure*}

\subsection{Key Components of Sharding} In a sharding system, increasing the throughput and capacity linearly while preserving security and decentralization is possible \cite{elastico}. However, this \textit{scale-out} mechanism or sharding poses new challenges to Blockchains. The main components and challenges for sharding protocols in public, permissionless blockchains that would impact their security and performance are as follows: 
\begin{enumerate}
    \item \textit{Identity establishment:} Prior to participation in the protocol, each node is required to establish a unique identity comprising elements like a public key, IP address, and a proof-of-work (PoW) solution. Based on this established identity, the node is then assigned to a suitable committee. Ensuring the distinctiveness of each node's identity is crucial for the system to prevent Sybil attacks \cite{sybil} effectively. However, this identity establishment process is not necessary in the context of a permissioned blockchain.
    \item \textit{Bootstrapping of committees:} To ensure security and mitigate potential attacks, it is essential to establish committees for the first epoch in an unbiased manner, maintaining an honest majority. This is crucial because any biases present in the initial setup can propagate through subsequent epochs, potentially leading to the failure of the system.
    \item \textit{Overlay setup for committees:} After the committees are established, each node initiates communication with fellow committee members to ascertain their identities. Within a blockchain, the committee structure forms a fully connected subgraph, encompassing all committee members. For this communication process, a gossip protocol \cite{gossip_protocol} is often employed, as it facilitates efficient information sharing among nodes. This setup is done within the consensus protocol, which is not discussed in detail in this paper. Interested readers are referred to the survey \cite{xiao2020survey} that is dedicated to blockchain consensus protocols to have a better understanding of this. 
    \item \textit{Intra-committee consensus:} Each committee member node employs a shared consensus protocol to reach an agreement on a unified set of transactions. During this phase, all honest members of the committee must concur on the proposed block.
    \item \textit{Cross-shard transactions:} Transactions spanning multiple zones or shards are termed cross-shard transactions. Sharding protocols must manage these transactions to ensure atomicity throughout the system. Typically, handling cross-shard transactions involves implementing strategies such as a locking mechanism \cite{omni}, employing relay transactions \cite{mono}, or breaking down the cross-shard transaction into several intra-shard transactions \cite{rapid}. These methods are crucial for achieving synchronization across the entire network.
    \item \textit{Epoch reconfiguration:} Given that sharding protocols operate in epochs, committees must be reconfigured after each epoch to maintain security, uphold an honest majority, and safeguard against adversarial influences. This reconfiguration relies on epoch-specific randomness, which is crucial for ensuring unbiased committee formation and preventing Sybil attacks \cite{sybil}.
    \item \textit{State sharding:} By definition, sharding protocols divide the whole network into smaller groups. Yet, to fully leverage the advantages of sharding, it is essential to implement state sharding. This involves dividing the entire database into smaller segments, each managed by a corresponding committee. The major challenge in state sharding is data migration overhead \cite{sschain}, which we will discuss in the discussion section (section VIII). 
\end{enumerate}
In the following sections, we discuss the core components of blockchain sharding, as described in Figure \ref{fig:components}.

\section{Identity Establishment and Committee Configuration}

\subsection{Identity Setup} Each participant, known as a node, must establish a unique identity in the blockchain network. This identity, typically comprising elements like a public key, is essential for maintaining the integrity and security of the network. In a decentralized network like blockchain, where there is no central authority to verify identities, establishing a secure method for identity establishment is essential. Blockchain relies on trust among nodes. By having a verified identity, each node can trust that others are also legitimate participants. This trust is vital for the network's integrity and for ensuring that transactions are legitimate and reliable. Unique identities prevent Sybil attacks \cite{sybil}, where a single entity creates multiple fake identities to gain undue influence in the network. By ensuring that each identity is unique and verifiable, the network can safeguard against such manipulative behaviors. 
Generally, the nodes need to find a PoW solution corresponding to their public key and IP address. The $epochRandomness$ generated in the previous epoch works as a $seed$ in the PoW, and it ensures that the PoW solution was not pre-calculated. The nodes need to find a hash value that satisfies the following: 
$$O = H(epochRandomness||IP||PK||nonce)\leq target$$

\subsection{Bootstrapping and Committee Configuration} 
Committees, comprised of groups of nodes, are assigned specific roles like transaction validation or block creation in blockchain sharding. The fundamental concept of sharding is to divide the network into multiple committees or shards, enabling nodes within each shard to operate independently and in parallel. This division aims to enhance efficiency and throughput Without periling the security of the blockchain. A crucial aspect of this process is ensuring that no adversary gains control over any single shard or multiple shards. Maintaining an honest majority in each shard is essential for this purpose. This objective is achieved by establishing committees that are both unbiased and resistant to Sybil attacks, where a single entity might try to control multiple identities. This careful formation of committees is key to preserving the integrity and security of the sharded blockchain network. 
 \subsubsection{Bootsrapping}
In RapidChain \cite{rapid}, they begin by creating a deterministic random graph, known as the sampler graph \cite{graph}. This graph facilitates the sampling of various groups, ensuring that the proportion of corrupted nodes within most groups closely aligns with their representation in the original set within a margin of $\delta$. During the bootstrapping phase of RapidChain, each participant in the bootstrapping protocol constructs this sampler graph locally. This construction uses a predefined, hard-coded seed and is based on the initial network size, which is common knowledge among all nodes, given the assumption that these nodes have already established their identities. This method ensures a controlled and predictable distribution of potentially corrupt nodes across the groups, thereby enhancing the security and integrity of the network right from its inception.\\ 
\textit{Sampler Graph:} 
In the sampler graph creation, a random bipartite graph $G(L,R)$ is created, and $d_R = O(\sqrt{n})$ where $d_R$ is the degree of each node in $R$. In the graph $G$, vertices of $L$ are the network nodes, and vertices in $R$ are the groups of the network. A node becomes a member of a group if it is connected within graph $G$. Consider $T$ as the largest subset of $L$ that contains faulty nodes and $S$ as any subset of groups within $R$. The event $\mathcal{E} (T, S)$ refers to the situation in which each group within set $S$ contains a greater number of edges connected to nodes in set $T$ than $\frac{|T|}{|L|} + \delta$. Intuitively, $\mathcal{E}$ represents the scenario where all groups in $S$ are ``bad,'' meaning that over of their $\frac{|T|}{|L|} + \delta$ members are faulty. It is proven that \cite{rapid_full} the likelihood of the event $\mathcal{E} (T, S)$ occurring is less than $2e^{(|L| + |R|)ln2-\delta^{2}d_R|S|/2}$. This formulation quantifies the probability, indicating that it remains significantly low under the defined parameters and constraints of the system. Besides, they select practical values for $|R|$ (the number of groups) and $d_R$ (the degree of connectivity within these groups) to ensure that the failure probability of the bootstrap phase is minimized. \\
Once the node groups have been established using the sampler graph, they participate in a randomized election procedure. However, before delving into the details of this procedure, it's important to explain how these groups can collectively reach a consensus on an impartial random number within a decentralized setting. \\ 
\textit{Subgroup Election:} Each group's members execute the DRG (Distributed Randomness Generation) protocol to generate a random string, denoted as \( s \) during the election phase. This string is then used to select representatives for the next level of groups. The process works as follows: each node, identified by its unique ID, calculates a hash value \( h \) using the formula \( h = H(s || ID) \), where \( H \) is a hash function treated as a random oracle. A node declares itself elected if its hash value \( h \) is less than or equal to \( 2^{256} - \nu \). After determining the elected nodes, all nodes in the group sign the pair (ID, s) of the $\nu$ nodes with the smallest hash values, \( h \). These signatures are then shared within the group, serving as verifiable proof of election for the chosen nodes. For practical implementation, the number of elected nodes per group, \( \nu \), is set to 2 in RapidChain. \\ 
\textit{Subgroup Peer Discovery:} Following the election of each subgroup, it is essential for all nodes to become aware of the identities of the elected nodes from every group. To facilitate this, the elected nodes disseminate their identity information along with proof to all other nodes. This proof comprises signatures from at least $d_R/2$ different group members on the pair $(ID, s)$. In instances where more than $\mathcal{E}$ nodes from a single group claim to be elected, it is an indication of dishonesty within that group. In such cases, all honest parties in the network will disregard any messages from the elected members of that suspicious group. \\ 
\textit{Committee Formation:} After the election protocol is carried out, the result is a group mainly made up of honest nodes, which is called the \textit{root group}. The root group's job is to pick members for the first shard, also known as the reference shard. Then, the reference committee divides all the nodes into different shards. This division is done randomly, but it's set up in a way that makes sure each committee has at least half of its members as honest nodes. \\ 
\textit{Election Network:} The election network is formed by linking together a series of sampler graphs, specifically $l$ of them, denoted as ${G(L_1, R_1), ..., G(L_l, R_l)}$. The specifications for all these sampler graphs are detailed in the protocol. Initially, the network’s $n$ nodes are part of $L_1$. Based on their connections in the graph, each node is assigned to various groups within $R_1$. Subsequently, every group conducts a subgroup election, following a specific protocol, to select a random subset of its members. The members who are elected in this stage then move on to become the nodes in $L_2$ for the next sampler graph $G(L_2, R_2)$. This sequence of elections and reassignments continues through to the final sampler graph $G(L_l, R_l)$. At this last stage, the process results in the formation of a single group termed the \textit{root} group. The structure of the election network is such that this leader group is likely to have a majority of honest members. 
The sharding protocols function within a permissionless environment, permitting unrestricted membership. In handling the initial randomness for bootstrapping the \textit{reference committee} or \textit{initial committee} in a permissionless environment, RapidChain follows this decentralized bootstrapping, which requires exchanging $O(n\sqrt{n})$ messages. At the same time, other sharding protocols initialize that common randomness by creating a genesis block with $O(n^2)$ messages.
\subsubsection{Committee Configuration}
In this section, we will discuss the committee configuration methods of sharding protocols. The protocols use an \textit{epoch randomness} to assign shards to the nodes, which is discussed in Section \textrm{VI}.  \\ 
\textbf{\textit{Elastico:}} In Elastico, like other sharding protocols, every node needs to solve a PoW puzzle to establish an identity to join the network. To ensure that the adversaries can not gain an advantage in advance to solve the puzzle, \textit{epoch randomness} is generated at the conclusion of the preceding epoch (discussed in section \textrm{VI}). The protocol allocates each identity to a committee randomly within \( 2^s \), which is identified using an \( s \)-bit committee identity. In particular, the final \( s \) bits of the identity determine the \( s \)-bit committee ID to which the processor is assigned. Every committee, defined by this \( s \)-bit ID, handles a distinct set of values. \\ 
\textbf{\textit{OmniLedger:}} After the validators finish a RandHound run successfully and the leader disseminates \( \text{rnd}_e \) along with a proof of its accuracy, all \( n \) duly enrolled validators are then able to authenticate and utilize \( \text{rnd}_e \). They compute a permutation \( \sigma_e \) of the sequence \( 1, \ldots, n \) and segment this permutation into \( m \) buckets of roughly equal size. This segmentation facilitates the distribution of nodes across various shards. \\ 
\textbf{\textit{RapidChain:}} RapidChain requires an off-chain PoW solution to establish identity and join the network by using generated epoch randomness. The nodes need to find $x$ to satisfy $O=H(T||PK||r_i||x) \leq target$, where $r_i$ is the epoch randomness. The nodes are subsequently distributed randomly among the committees according to generated epoch randomness $r_i$. \\ 
\textbf{\textit{Monoxide:}} In Monoxide, the user's address, which is its public key's hash value, is partitioned uniformly into $2^s$ zones, where the first $s$ bits is the zone index. This approach is similar to that of Elastico. \\ 
\textbf{\textit{TEE-based Protocol:}} Forming shards securely necessitates the use of an impartial random number, \( \text{rnd} \), as the foundational seed for allocating nodes to committees. With \( \text{rnd} \) provided, nodes determine their respective committee placements by generating a random permutation \( \pi \) of the series [1 : N], using \( \text{rnd} \) as the seed. This permutation \( \pi \) is then segmented into chunks of nearly equal size, with each segment corresponding to the members of a particular committee. \\ 
\textbf{\textit{Pyramid:}} In $Pyramid,$ the nodes also need to solve the PoW puzzle based on its public key and epoch randomness. The nodes are then assigned to \textit{i-shards} or \textit{b-shards} according to their identity and epoch randomness, where \textit{i-shards} verify internal transactions and \textit{b-shards} process the cross-shard transactions. \\ 
\textbf{\textit{Brokerchain:}} $BrokerChain$ also requires solving a PoW puzzle for the nodes to establish their identity, and then they are assigned to shards according to the last few bits of the solution. Brokerchain consists of two types of shards: \textit{M-shards} for generating TX blocks and \textit{P-shard} for partitioning account states to reduce the number of cross-shard transactions. 

\section{Intra-Committee Consensus} 
The primary objective of a blockchain consensus protocol is to guarantee that all nodes participating in the network reach an agreement on a single, unified transaction history. This agreed-upon history is chronologically organized and recorded as a blockchain. In the sharding mechanisms, the intra-committee consensus ensures that the nodes in the same shard agree upon a single value, transaction, or block header and that the honest nodes can discard any invalid transaction or malicious messages to ensure the security of the shard. According to \cite{xiao2020survey}, the requirements of blockchain consensus protocol are: 
\begin{itemize}
    \item \textit{Termination:} Consensus makes sure that an honest node in the network either accepts a new valid transaction or discards invalids. 
    \item \textit{Agreement:} Each new transaction and its corresponding block must be uniformly accepted or rejected by all honest nodes. Furthermore, every honest node should assign the same sequence number to any block that is accepted.
    \item \textit{Validity:} When a valid transaction or block is received identically by all nodes, it should be included in the ledger.
    \item \textit{Integrity:} All transactions accepted by the honest nodes must be consistent, ensuring no double spending occurs. Additionally, all accepted blocks need to be properly created and linked in a hash chain, following a chronological sequence.
\end{itemize}
Consensus mechanisms in sharding protocols serve mainly two purposes. The first is establishing a Sybil-resistant and valid identity, as discussed in section \textrm{III}. The other one is to reach an agreement within each shard and the whole network. According to this study \cite{sok}, blockchain consensus can be categorized into two groups. Proof-of-stake, proof-of-work, or other ``proof-of-something" consensus protocols can be categorized as \textit{PoX}. Another category is \textit{BFT} based consensus mechanisms. 
\subsection{PoX}
Sharding protocols require PoX-based protocols to establish valid identity for the nodes. To join the sharding network, the nodes must verify that they are valid nodes and not conduct any Sybil attack. In order to do this, they need to show some effort, which requires computational resources, tokens, or other stakes or resources. 

\textit{Proof-of-Work:} The first cryptocurrency,  Bitcoin \cite{bitcoin}, introduced proof-of-work in blockchain, which is used as the identity establishment mechanism in blockchain sharding protocols, also known as Nakamoto consensus. In proof-of-work, the nodes that \textit{propose} a block to the network must solve a cryptographic puzzle. The solution to that puzzle must meet the requirements for the proposed block to be considered valid and to be added to the chain. To solve that puzzle, the \textit{miner} needs to use his computational resources to find a hash value that meets the requirement. Similarly, in sharding protocols, the nodes that want to join the network must solve the hash puzzle to meet the target value and to establish themselves as valid nodes, as discussed in section \textrm{III}. The Nakamoto consensus follows the following criteria \cite{xiao2020survey}: 
\begin{itemize}
    \item \textit{Proof of Work:} Miners need to find a hash value that meets certain requirements to validate a block and to add this to the chain. The difficulty or target to solve the hash puzzle is dynamically adjusted to maintain an appropriate interval for block creation. 
    \item \textit{Gossiping Rule:} When a node receives a new transaction or block, it broadcasts to its peers immediately. 
    \item \textit{Validation Rule:} Before broadcasting the transaction or block, it must be validated by the message sender. For transactions, it checks for double-spending, and for block headers, it checks for a valid solution of the hash puzzle. 
    \item \textit{Longest-Chain Rule:} The longest chain confirms that the computational power used for this chain is more than any other existing forks \cite{fork}. So, when there are multiple forks, the miners must choose the longest chain and should work on appending it. 
    \item \textit{Incentive Mechanisms:} There are two kinds of incentives in this consensus protocol. There is a fixed block generation fee known as the \textit{block reward} or the \textit{coinbase} transaction, which is halved every four years. Besides, the miners get the \textit{transaction fees} collected from all the transactions included in the generated block. 
\end{itemize}
In blockchain sharding systems, the nodes need to solve a PoW hash puzzle to establish identity and join the network. As it requires heavy computation and real hardware-based resources, it is highly unlikely to create or replicate as many nodes as possible to join the network in order to manipulate any specific shard or multiple shards, which is known as the Sybil attack \cite{sybil}. \\
PoW is generally prone to some attacks, such as double spending attack \cite{double_spending}, 51\% attack \cite{attacks}, TCP vulnerabilities attack \cite{bijack} etc. Several approaches have been proposed, such as GHOST \cite{ghost} protocol, Bitcoin-NG \cite{ng}, etc., to improve the performance and security of the original PoW. However, since most sharding protocols mostly use PoW as identity establishment protocol, not to add or discard any block to the chain, the network is not susceptible to these PoW-based attacks. 

\textit{Proof-of-Stake:} Solving the hash puzzle in PoW requires huge computational resources and power. To minimize the use or waste of physical resources, such as GPU, proof-of-stake(PoS) requires virtual resources, such as tokens, to propose or vote on new blocks in the chain. Unlike PoW, there is no mining in PoS, so the miners or nodes that propose new blocks are called the \textit{validators} or \textit{minters}. When the validators try to manipulate the network to gain advantages or conduct a Sybil attack, there is a risk of losing the stake that is invested in the chain. That is why PoS is more secure in such cases than PoW. There are four classes of PoS: \\
\textbf{\textit{Chain-Based PoS:}} It is usually similar to PoW, except for the fact that the block generation in this method follows PoS instead of PoW. \textit{Peercoin} \cite{peercoin} and \textit{Nxt} \cite{nxt} follow chain-based PoS. Since it is the adaptation of PoW, it can also tolerate up to 50\% malicious stakes. \\
\textbf{\textit{Committee-Based PoS:}} Committee-based PoS uses a more structured approach by forming a committee of stakeholders, chosen based on their stakes, to take turns generating blocks. This selection relies on a secure \textit{multiparty computation} (MPC) \cite{mpc} scheme, a type of distributed computing where multiple parties, starting with individual inputs, arrive at a uniform output. The committee selection can be made privacy-preserving and verifiable through a verifiable random function (VRF) \cite{vrf}, allowing only the selected stakeholders to know of their committee inclusion. \textit{Chain of activity} \cite{coa}, \textit{Ouroboros} \cite{ouroboros}, \textit{Ouroboro Praos} \cite{ouro_praos}, \textit{Snow White} \cite{snow_white}, etc., are some committe-based PoS schemes. However, committee-based PoS also follows the longest-chain rule of PoW. \\
\textbf{\textit{BFT-based PoS:}} BFT-based PoS, also known as hybrid PoS-BFT, leverages deterministic block finalization instead of probabilistic PoW-adapted chain-based PoS or committee-based PoS. In this scheme, the most-recent-stable-checkpoint replaces the PoW-based schemes' longest-chain rule. \textit{Tendermint} \cite{tendermint}, \textit{Algorand} \cite{algorand}, ByzCoin \cite{byzcoin} are some well-known BFT-based PoS schemes. Generally, BFT-based PoS consensus mechanisms can tolerate up to $1/3$ of Byzantine validators. \\
\textbf{\textit{Delegated PoS (DPoS):}} To mitigate the communication overhead of consensus protocols, DPoS goes through a process called \textit{delegaraion} process to elect the consensus group called \textit{delegates}. This group consists of a fixed number of members who run the consensus on behalf of all the stakeholders. \textit{BitShares 2.0} \cite{bitshares}, \textit{Cosmos} \cite{cosmos}, \textit{EOSIO} \cite{eos} are some schemes that are based on BFT-based PoS. This consensus mechanism can also tolerate up to $1/3$ of byzantine validators since it is based on BFT consensus. 

Non-BFT-based PoS consensus mechanisms possess the vulnerability of \textit{costless simulation}, which refers to a player's ability to recreate any part of the blockchain's history without expending real effort or resources. This is possible in PoS systems as they do not require intensive computation. PoS also faces the same \textit{centralization} risk as PoW-based schemes. When more than 50\% of the stakes are owned by any specific validator or stakeholder, regardless of the validator's nature, the whole system will be dominated by him. Since one of the major goals of the sharding systems is to ensure decentralization of the network, non-BFT-based PoS schemes are usually not used in blockchain sharding. However, there are some countermeasures to tackle this centralization of the network \cite{xiao2020survey}.

Besides these two major consensus mechanisms of PoX, some other consensus protocols exist where \textit{miners} or \textit{validators} need to prove the ownership of tokens in the network. Some of them are: \textit{proof-of-burn} \cite{pob}, \textit{proof-of-activity} \cite{poa}, \textit{proof-of-elapsed time} \cite{poet}, \textit{proof-of-coin age} \cite{peercoin}. For detailed information about these protocols, readers are encouraged to read the respective papers.

\subsection{BFT}
Blockchain sharding protocols mostly use BFT as the intra-shard consensus mechanism, where the shards select a leader to agree on valid block headers or to reach a consensus regarding the ledger's state. In this part, we will briefly discuss the BFT protocols, their system security, and the challenges the sharding protocols face that use BFT-based consensus mechanisms.

BFT also follows the four requirements of blockchain consensus \cite{bftcon,bftcon2}: \textit{Termination, Agreement, Validity, and Integrity}. Let $f$ be the fraction of faulty nodes in the Byzantine system. So, to meet the consensus requirements, the network needs to satisfy the condition: $N\geq3f+1$ \cite{bftproof}. \\ 
\textbf{\textit{PBFT:}} In the blockchain system, typically, the BFT protocol means PBFT consensus protocol, which was developed in 1999 \cite{pbft}. PBFT protocol operates in three phases: \textit{pre-prepare, prepare,} and \textit{commit}. The \textit{leader} orders the sequence of the client's requests and sends them to other nodes. All the nodes then complete the three-phase protocol to reach the consensus. Here, the safety of the process depends on the leader only, so the protocol can maintain stability even if there are timing violations by the other nodes. If the other nodes detect a faulty leader, the \textit{view-change} sub-protocol is triggered, and a new leader is selected. This leader-based consensus protocol works well for the sharding systems where the shards or the committees mainly rely on honest leaders for each epoch. However, leader-based sharding systems face challenges in maintaining honest leadership and face several consequences when a malicious leader controls any specific shard.

Besides this leader-based BFT protocol, some leaderless BFT protocols are also used in blockchain systems, mainly asynchronous systems. We will now briefly discuss these protocols.

\textbf{\textit{HoneyBadger:}} HoneyBadger \cite{honeybadger} is the first BFT-based protocol designed for asynchronous blockchain systems. This protocol combines erasure-coded RBC \cite{rbc} and common-coin-based ABA \cite{aba} to construct ACS construction \cite{acs}. Using threshold public key encryption (TPKE) \cite{tpke}, HoneyBadger performs consensus in ciphertexts to prevent malicious parties from manipulating transactions or the network. However, for using TPKE, HoneBadger possesses a higher cryptographic overhead than partially synchronous BFT protocols like PBFT. As HoneyBadger works in a leaderless system with an asynchronous network, it does not suffer from the bottlenecks of leader-based BFT protocols, like bandwidth limitation of the leader, selecting a new leader in each rotation, etc. On the other hand, transaction confirmation latency is higher in this asynchronous protocol as there is no fixed delay in message communication. Addressing this issue, BEAT \cite{beat} outperforms HoneyBadger by using another threshold encryption mechanism \cite{beat_threshold}. 

Now we will discuss the public blockchain sharding schemes' consensus protocols and how they reach the intra-shard consensus. 
\subsection{Current Sharding Protocols' Consensus Mechanism}
\textbf{\textit{Elastico:}} Any authenticated Byzantine agreement protocol, such as PolyByz \cite{polybyz} or PBFT \cite{pbft}, can be employed in Elastico for intra-shard consensus on a set of transactions. Once consensus is reached, the agreed-upon transaction set must be validated with at least \(c/2 + 1\) signatures, guaranteeing that a minimum of one honest member has reviewed and approved the transactions. After this, each committee member forwards the signed transactions and the accompanying signatures to the final committee. The final committee members are identified by obtaining the list of final committee members from the directory once more. The final committee then confirms the selected transactions by verifying they have the required signatures. The final digest can be a Merkle root hash representing all the values. \\
\textit{\textbf{OmniLedger:}} ByzCoin integrates Proof of Work (PoW) with Byzantine Fault Tolerance (BFT) algorithms in a hierarchical, tree-based structure, utilizing scalable collective signing (CoSi) \cite{cosi} to achieve this combination. The problem with ByzCoin is that despite the scalability of ByzCoin's consensus process, the total processing capacity of the system remains unchanged, regardless of the number of participants. OmniLedger introduces a new consensus mechanism called ByzCoinX \cite{omni}, which trades off some scalability of the original ByzCoin for robustness by a two-level tree structure within the consensus group. In ByzCoinX, the protocol leader randomly selects a \textit{group leader} at the beginning of each epoch. The group leaders are responsible for managing communication to achieve consensus within their respective shards. They must communicate within a set timeout period; if they fail to do so, the protocol leader appoints a new group leader to facilitate the consensus process. A leader must secure votes from two-thirds of the participants to move to the next consensus phase. Additionally, ByzCoinX employs a view-change window, similar to the PBFT protocol, to replace any leader that is deemed faulty. With this algorithm, the problem of BFT-based 1\% attacks in sharding is solved by increasing the size of the shard to include hundreds or even thousands of participants. \\ 
\textbf{\textit{RapidChain:}} RapidChain's BFT-based consensus protocol consists of two main parts. The first is a \textit{gossiping protocol} where the nodes within each shard propagate transactions or block headers. Another is a \textit{synchronous protocol} used to reach an agreement on the block header. 
\subsubsection{Gossiping Protocol} RapidChain adapts \textit{Information Dispersal Algorithm (IDA)} \cite{ida} to establish the \textit{IDA-Gossip} protocol for message communication within the shards. In this protocol, in a network shard with $\phi$ malicious or faulty nodes, a message $M$ is divided into $(1-\phi)k$-equal sized segments or chunks, which are $M_1,M_2,...,M_{(1-\phi)k}$. To ensure the integrity of the message where $\phi$ faulty nodes are present, RapidChain integrates an erasure code scheme \cite{erasure} so that the original message can be constructed using any set of $(1-\phi)k$ chunks. The IDA-Gossip protocol does not function as a reliable broadcast protocol because it does not prevent the sender from equivocating. However, in comparison to reliable broadcast protocols \cite{async}, IDA-Gossip demands significantly less communication overhead and offers quicker propagation, especially for large blocks of transactions, such as the approximately 2MB blocks found in RapidChain. 
\subsubsection{Synchronous Consensus} RapidChain employs a modified version of the synchronous consensus protocol developed by Ren et al. \cite{sync_rapid}. This adaptation enables the attainment of an optimal resilience level of \( f < 1/2 \) in committees, allowing for smaller committee sizes while still achieving a higher total resilience of $1/3$, an improvement over previous sharding protocols \cite{elastico,omni}. In RapidChain, the synchronous consensus protocol is utilized specifically to attain consensus on a digest of the block put forward by a committee member. Consequently, the remainder of the protocol operates effectively over partially synchronous channels, utilizing optimistic timeouts to ensure responsiveness, similar to \textit{Elastico}. A problem with this synchronous consensus protocol is that when an output committee leader is malicious, he may send a deceiving message to the input committee regarding a transaction that all honest nodes in the shard have not accepted. This problem occurs as synchronous protocols are \textit{round-driven} instead of \textit{event-driven} like asynchronous protocols.

\begin{figure}[h]
    \centering
    \includegraphics[width=\columnwidth]{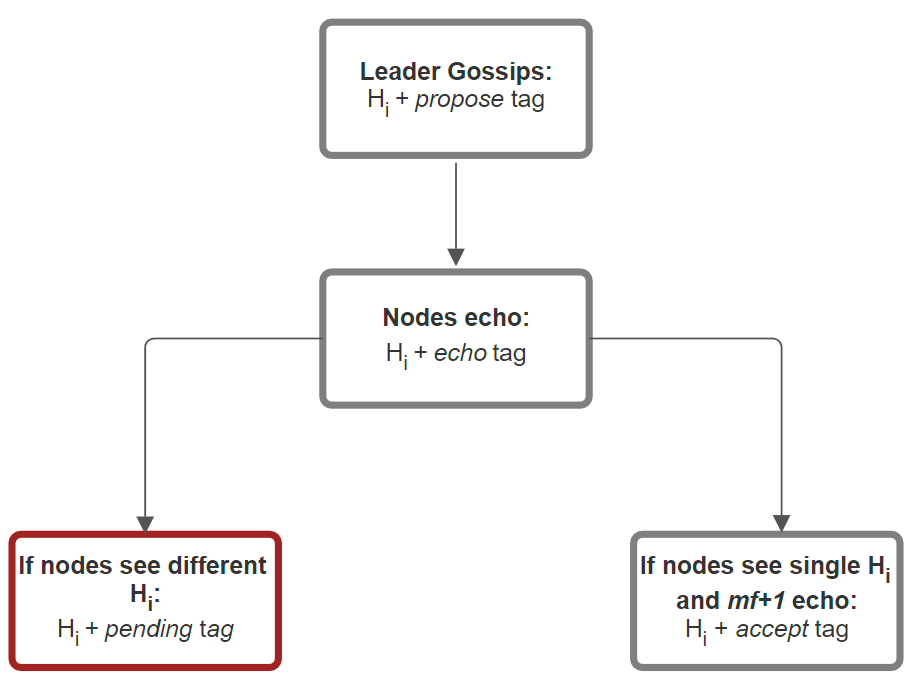}
    \caption{RapidChain consensus rounds}
    \label{fig:Rapid}
\end{figure}

\textit{Protocol Details:} At each iteration in RapidChain, a committee randomly selects a leader using epoch randomness to drive the consensus protocol. The leader compiles all received transactions into a block \( B_i \) and uses IDA-Gossip to distribute it, creating a block header \( H_i \) that includes the iteration number and the Merkle tree root from IDA-Gossip. The leader then initiates a consensus protocol on \( H_i \). The consensus protocol involves four synchronous rounds (Figure \ref{fig:Rapid}). First, the leader sends a message with \( H_i \) and a \textit{propose} tag. Second, other nodes echo this header by re-gossiping \( H_i \) with an \textit{echo} tag, ensuring all honest nodes see any header versions circulated. If the leader sends out multiple message versions, indicating equivocation, this will be detected. In the third round, if an honest node receives differing headers for the same iteration, it identifies the leader as corrupt and gossips a modified header \( H'_i \) with a \textit{pending} tag, where \( H'_i \) contains a null Merkle root and the iteration number. If an honest node in the shard receives \( mf + 1 \) echoes of a singular and consistent header \( H_i \) for a given iteration \( i \), it acknowledges \( H_i \) as valid. Subsequently, it broadcasts \( H_i \) along with all \( mf + 1 \) echoes of \( H_i \) using an \textit{accept} tag. These \( mf + 1 \) echoes act as evidence supporting the node's acceptance of \( H_i \). It's noteworthy that creating this proof is impossible unless the leader has initially shared \( H_i \) with at least one honest node. When an honest node accepts a header, it ensures that all other honest nodes will also accept that same header or will completely reject any header from the leader. In situations where the leader is corrupt, some honest nodes will disapprove of the header, marking it as \textit{pending}. To maximize throughput, RapidChain allows the committee leader to re-propose the pending block headers. The \textit{pending} or \textit{accepted} votes of the nodes can be either \textit{temporary} or \textit{permanent} relative to the current iteration. When a node accepts a header, it will broadcast the \textit{accept} tag and the header. So all the nodes then know that node's vote and that node will not broadcast any more headers for the iteration. This process makes it impossible for malicious leaders to perform denial-of-service attacks by forcing the honest nodes to echo many non-pending block headers.\\
\textbf{\textit{Monoxide:}} Monoxide is the first sharding protocol that relies on PoW for intra-shard consensus. Like the Bitcoin network, in Monoxide, the miner solves a hash puzzle to add blocks to the chain. The rationale behind using PoW as the consensus mechanism is enabling \textit{Chu-ko-nu} mining. \\ 
\textit{Chu-ko-nu Mining:} Chu-ko-nu mining, inspired by \cite{chukonu_insp}, enables a miner to utilize a single Proof of Work (PoW) solution to simultaneously generate multiple blocks across different zones, with the limitation that only one block can be created per zone. Miners are allowed to create blocks based on their computational capacities. Without Chu-ko-nu mining, a miner solves the hash puzzle as in Bitcoin or other PoW-based networks where he needs to find a hash value less than the target as follows: 
$$h(H_i,\eta_i)<\tau$$
where $\eta$ is the nonce required to find the hash of block header $H_i$, which is less than the target $\tau$. With Chu-ko-nu mining, several shards comprise the Merkle Patricia Tree (MPT) \cite{ethereum} root containing all proposed blocks in this mining. The miner needs to find a nonce $\eta$ that satisfies the following: $$h(\eta||h(H_0||MPT_M))\leq \tau$$
where $\tau$ is the target difficulty, $H_0$ is the chaining-header of the blocks, $MPT_M$ is the MPT consisting of all proposed blocks. \\ 
Chu-ko-nu mining enhances security by making an attack on a single zone as challenging as an attack on the entire network. If $m_p$ is the total physical hash rate of Chu-ko-nu miners, $m_d$ is the total physical hash rate of non-Chu-ko-nu miners in a sharding network having $2^s$ shards, then the effective hash rate $m_s$ is: 
$$m_s=\frac{m_d}{2^s}+m_p$$
So, an adversary can control the network if it can obtain $>\frac{m_s}{2}$ hash rate. Since currently, PoW-based minings are run by mining pools, it will be extremely hard for the adversary to obtain such higher hashing power to manipulate the network. So, to get the maximum rewards, honest miners will try to participate in all shards with Chu-ko-nu mining, which will amplify the effective mining power and prevent adversaries from targeting any shard to gain control. \\ 
\textbf{\textit{TEE-Based Sharding:}} This TEE-based sharding protocol \cite{towards} adopts a protocol \cite{towards_proto} in order to prevent byzantine nodes from equivocating in the network. If equivocating can be prevented, the system can tolerate as many as \( f = \frac{N-1}{2} \) non-equivocating Byzantine failures in a network comprising $N$ nodes. In the adopted protocol, the trusted log abstraction, which is used to prevent equivocation, is securely stored within the Trusted Execution Environment (TEE) to prevent tampering by attackers. They also introduce an improved proof-of-elapsed-time (PoET) to restrict the competition of nodes to propose the next block. By doing so, the \textit{stale block} rate is also lessened \cite{stale}. \\
\textbf{\textit{Pyramid:}} Layered sharding protocol \textit{Pyramid} consists of two types of shards: \textit{i-shards} for handling intra-shard transactions and \textit{b-shards} for handling cross-shard transactions. Both kinds of shards run a BFT protocol, similar to ByzCoinX in OmliLedger, to achieve consensus. \\ 
\textbf{\textit{RepChain:}} RepChain is a sharding scheme based on reputation. Within each shard, it establishes two chains: a transaction chain (TB) using the Raft consensus mechanism, and a reputation chain (RB) using Byzantine Fault Tolerance (BFT), specifically CSBFT. For intra-shard consensus, the leader shares the transactions list \textit{(TxList)} with all validators. Each validator can send \textit{accpet, reject} or \textit{unknown} decision for each transaction in the list. Then, the leader decides which transactions to add to the chain. The validators send \textit{warning} message to other validators if they suspect the leader is malicious, and then the \textit{view-change} is triggered as in typical BFT protocols. \\ 
\textbf{\textit{BrokerChain:}} BrokerChain consists of multiple \textit{M-shards} and a \textit{P-shard}. \textit{M-shards} run a PBFT-based intra-shard consensus protocol at the beginning of each epoch to generate transaction blocks.

\section{Cross-Shard Transactions} 
The main objective of blockchain sharding protocols is to achieve higher throughput while ensuring the security of the system. If a single shard network is overflowed with abundant transactions, the throughput will not increase, and the network of that shard will be congested with communication messages, which will result in longer transaction confirmation latency, and thus, the performance of the sharding system will deteriorate. To efficiently use the shards in the network, transactions must be distributed among multiple shards. In blockchain, a transaction consists of multiple inputs and outputs. A transaction is considered a \textit{cross-shard transaction} if the inputs and outputs of the transaction involve more than one shard. As the transactions are randomly distributed in the network, most of the transactions will be cross-shard, which means, the input shards and output shards will be different. The previous protocols show that the number of cross-shard transactions in a \textit{UTXO} (Unspent Transaction Output) based sharding system is more than 90\% \cite{omni}, and up to 90\% for \textit{account-balance} transaction model \cite{mono}. Taking that into consideration, the sharding protocols need to handle cross-shard transactions properly to meet the following requirements: 
\begin{itemize}
    \item \textit{Atomicity:} These transactions must be atomic; either the entire transaction is successfully processed across all involved shards, or none of it is. Improper handling could lead to partial updates, double-spending \cite{doublespending}, or other potential attacks \cite{attacks}.
    \item \textit{Integrity:} Each shard maintains a part of the blockchain's state. Cross-shard transactions modify states across multiple shards. A proper transaction handling mechanism is necessary to ensure the integrity of the ledgers. 
    \item \textit{Network Security:} Handling cross-shard transactions involves complex coordination and communication between shards. If not done securely, this can become a vulnerability point for attacks, such as replay attacks \cite{replay}, where the same transaction is maliciously repeated across shards. 
    \item \textit{Performance and Scalability:} One of the primary goals of sharding is to improve the blockchain's scalability and performance. Efficient handling of cross-shard transactions is crucial to achieving this. Poor management can lead to bottlenecks, negating the benefits of sharding by causing delays and reducing throughput.
\end{itemize}
In this section, we will discuss the transaction models of blockchain sharding systems, the mechanisms to perform the cross-shard transactions, and the vulnerabilities or limitations of the state-of-the-art sharding protocols in handling these transactions. 
\subsection{Transaction Model}
\textit{UTXO Model:} The most widely used transaction model in blockchain systems is the UTXO model, which Bitcoin first used. Most of the sharding protocols also follow this model, including Elastico, OmniLedger, RapidChain, etc., In this model, a transaction consists of multiple inputs and outputs, where the inputs are unspent outputs of previous transactions. The inputs are marked as the spent outputs and removed from the UTXO transaction pool, where only unspent outputs are stored. In the sharding systems, a transaction containing multiple inputs and outputs may involve multiple shards. Both types of transactions (intra and cross) handling mechanisms are important to handle such transactions. The intra-shard transactions can be handled easily since all the nodes in the shard share the same ledger or state. But to handle the cross-shard transactions, the system must ensure that the different involving shards are either \textit{accpeting} or \textit{discarding} the same transaction together to prevent double-spending or exploitation of other vulnerabilities. \\
\textit{Account-Balance Model:} The Account-balance model shares the same idea as the bank account model in general. Unlike the UTXO model, the accounts in this model are not destroyed or consumed by other accounts. The balance of the accounts increases or decreases depending on the input or output of the transactions. One major benefit of this account-balance model over the UTXO model is the space savings. This model does not require multiple inputs and multiple outputs to refer to one single entity. Ethereum \cite{ethereum}, as well as many other cryptocurrencies \cite{mono,skychain}, follow this account-balance transaction model.
\subsection{Cross-shard Transaction Handling Mechanisms}
When a client or user initiates a transaction involving multiple shards, the transaction is routed and then executed upon verification depending on the routing protocol and cross-shard transaction handling mechanism of that protocol. In this section, we will discuss the cross-shard transaction handling mechanisms of sharding protocols and how they obtain the atomicity of the transactions. 
\subsubsection{OmniLedger - Atomix} 
OmniLedger handles the cross-shard transactions and obtains atomicity via a \textit{lock-unlock} based mechanism called \textit{Atomix} \cite{omni}. This mechanism consists of the following three phases: \\
Firstly, a cross-shard transaction is created from the client's end, which is then gossiped through the network, and eventually, the involved intra-shards get the transaction. \\ 
After validating the own shard's input transaction, if the transaction is valid, the input shards \textit{lock} the corresponding input UTXOs and send \textit{proof-of-acceptance}. Otherwise, it sends \textit{proof-of-rejection}. The client eventually gets all \textit{proof-of-acceptance} it needs or one or more \textit{proof-of-rejection}. \\ 
Lastly, if the client gets all the \textit{proof-of-acceptance}, it gossips \textit{unlock-to-commit} transaction, including the proof. The involved output shards then validate the output transactions and add them to the ledger. 
If the client gets any \textit{proof-of-rejection} in the previous step, then it sends the \textit{unlock-to-abort} transaction to the input shards with the proof. After getting the proof and request to abort, the input shards make the UTXO available to spend again. \\ 
The \textit{Atomix} protocol, however, does not specify whether the client has any specific routing algorithm to gossip the transaction to the input and output shards. If the client needs to gossip the transaction to the entire network in order to reach the input and output shards, it will incur a large communication overhead. Besides, the client needs to gather the proof and send it to the output shards or, again, the input shards (if rejected), which might be burdensome for lightweight nodes.  
\subsubsection{RapidChain} RapidChain handles the cross-shard transactions by splitting them and creating multiple transactions to make the whole transaction multiple 'single input-single output' transactions. It is easier for both the input and output shards to verify and add the transactions to the block in this way. In this mechanism, Let $I_1$ and $I_2$ be the two inputs of a transaction that belong to input shards $S_1$ and $S_2$ (Figure \ref{fig:rapid_tx}, and $O$ is the output that belongs to output shard $S_3$. When the client gossips the transaction to the network, it eventually reaches the output shard, which handles the cross-shard transactions. The output shard $S_3$ creates two separate transactions where the inputs are $I_1$ and $I_2$, and the outputs are $I'_1$ and $I'_2$. $S_3$ then sends these transactions to $S_1$ and $S_2$ to verify, who then send back $I'_1$ and $I'_2$ to $S_3$ upon verification. Finally, $S_3$ creates another transaction consisting of the inputs $I'_1$ and $I'_2$, and output $O$. 
\begin{figure}
    \centering
    \includegraphics[width=\columnwidth]{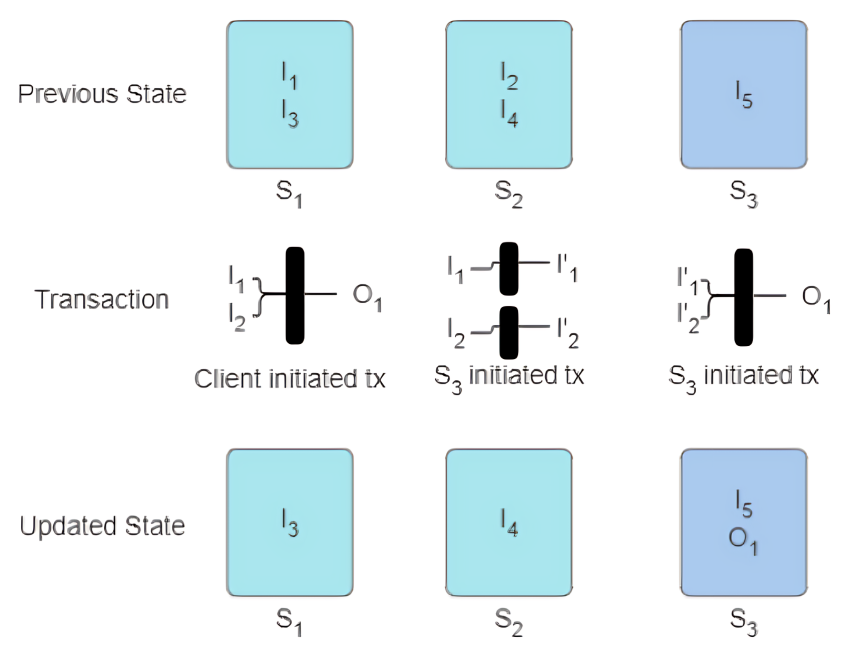}
    \caption{Cross-Shard Tx in RapidChain}
    \label{fig:rapid_tx}
\end{figure}
To reduce the communication overhead, RapidChain uses the idea of \textit{Kademlia} routing protocol \cite{kademlia}. This routing protocol uses a metric of distance, such as Hamming distance \cite{hamming}, to efficiently propagate messages from one shard to another. For each cross-shard transaction, RapidChain creates $x+1$ additional transactions where $x$ is the number of different input shards in the transaction. The input shards do not send any proof to the output shard whether they have added the input transaction to their ledger or not. This arises \textit{double spending} problem if the input shard's leader is malicious. To solve this problem, Merkle root \cite{merkle} can be included with the transaction to prove that the transaction is added to the input shard's ledger. Another limitation of this handling mechanism is that it can not handle cross-shard transactions involving multiple output shards. Since the cross-shard transactions are routed to the output shards using the shard index, it is possible to perform a denial-of-service attack by flooding the network with invalid transactions using the specific shard index.
\subsubsection{Monoxide - Eventual Atomicity} Monoxide uses \textit{relay transaction} to handle cross-shard transactions and to eventually get atomicity. In this mechanism, the input shard validates and adds the input transaction to its ledger and sends it to the output shard by using \textit{dynamic hash table} (DHT) \cite{dht} (Figure \ref{fig:mono}). Using DHT to route the transaction to the output shard reduces the communication overhead as it is sent directly to the output shard using the \textit{shard index}. The output shard then adds the transaction to its ledger if the verification is valid. The problem with this mechanism is that if the transaction fee is comparatively low, then no miner in the output shard might want to add it to their block, even if they are honest. This leads to partial completion of the transaction. Another problem with this mechanism is it can not handle transactions with multiple inputs. 
\begin{figure}
    \centering
    \includegraphics[width=\columnwidth]{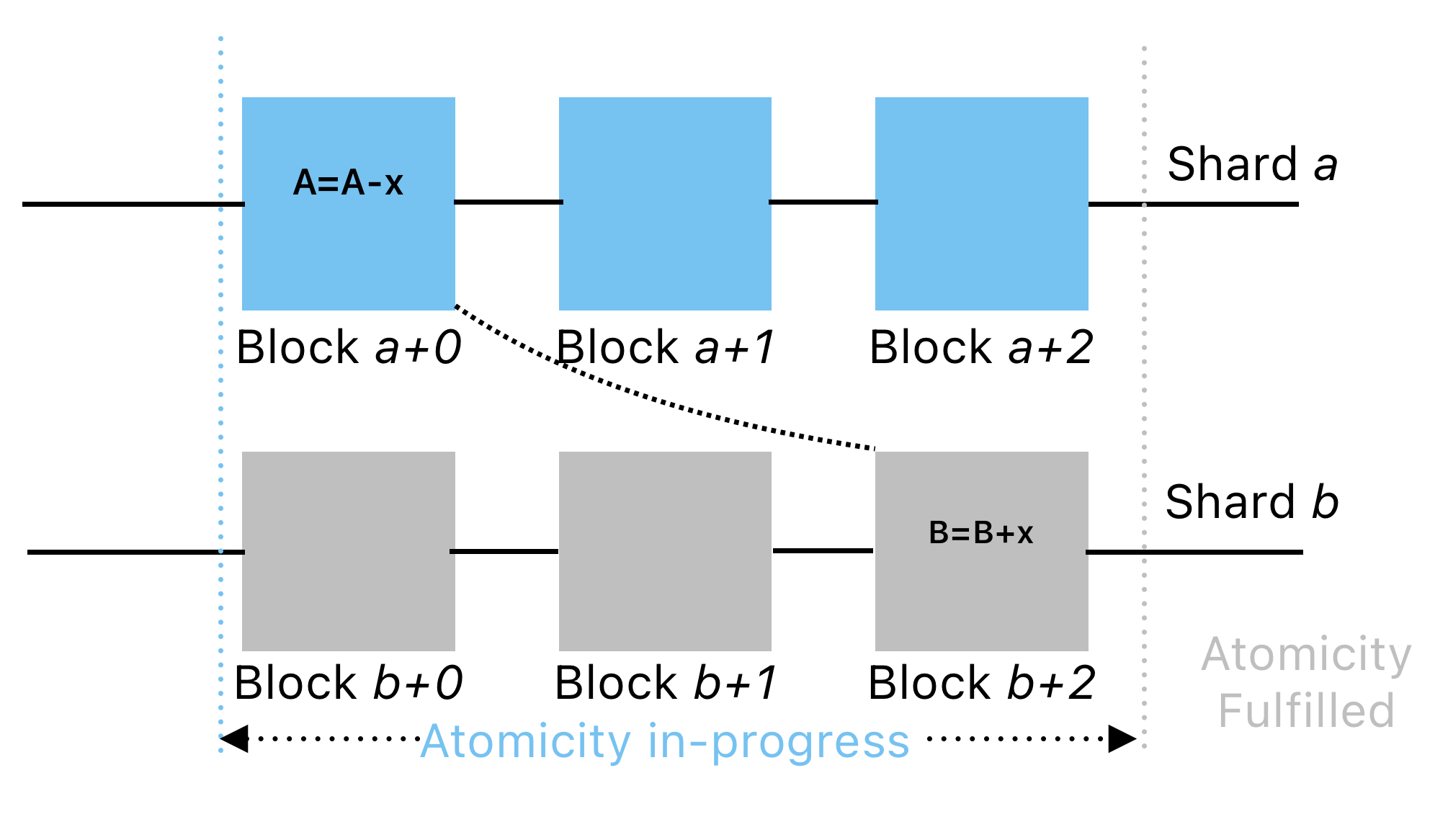}
    \caption{Relay transaction in Monoxide}
    \label{fig:mono}
\end{figure}
\subsubsection{TEE-based Sharding} The TEE-based sharding protocol uses a 2-phase commit and 2-phase lock protocol, which is similar to OmniLedger's \textit{lock-unlock} protocol. It, however, suffers from the same communication overhead problem as OmniLedger. 
\subsubsection{Pyramid} 
In Pyramid, \textit{b-shards} handle the cross-shard transactions. The nodes in \textit{b-shards} are also the nodes of multiple \textit{i-shards}. So, each cross-shard transaction is an intra-shard transaction for \textit{b-shards}. This mechanism has three phases: \textit{pre-prepare, prepare and commit}. In the \textit{pre-prepare} phase, the \textit{b-shard} reaches a consensus for the transaction and then sends it to corresponding \textit{i-shards} to update their local state. After verifying the transactions, the corresponding \textit{i-shards} update their local state and send \textit{accept} message to \textit{b-shard} in the \textit{prepare} phase. In the \textit{commit} phase, \textit{b-shard} finally updates its local shard when it receives all the \textit{accept} messages from all \textit{i-shards}.
\subsubsection{Brokerchain}
\textit{Brokerchain} processes the cross-shard transactions with \textit{broker} accounts. In this protocol, accounts can be segmented and assigned to multiple shards to enable cross-transaction with the help of these broker accounts. The \textit{broker} accounts act as a mediator to split the cross-shard transactions into multiple intra-shard transactions. Since the broker's account is segmented and is in multiple shards, it can treat the cross-shard transactions as multiple intra-shard transactions.   

\section{Epoch Randomness and Committee Reconfiguration} 
Blockchain sharding protocols divide the whole network among many shards or zones, and the nodes are assigned to specific zones where they run the intra-consensus protocol and take part in cross-shard transactions. If the assignment to shards is unfair, the shards will be susceptible to attacks, and eventually, the system will fail. If an adversary can deliberately be assigned to a shard of its choice, then it might manipulate the shard and act maliciously to interrupt the operations of the shard and the whole network. The assignment must maintain the honest majority of the network and the specific threshold of the faulty nodes. To achieve this, the nodes must be assigned to the shards \textit{randomly}. If the committee assignments are \textit{static}, the adversaries can leave and join the network to be assigned to their desired shard, where they can gradually increase the number of malicious nodes to reach the threshold of faulty nodes in the consensus mechanism, and, eventually, can manipulate the network. To solve this issue, committees must be reconfigured after a certain period of time. In the blockchain sharding system, this certain period of time is called \textit{epoch}. The sharding system works in epochs, and the committees are reconfigured after each epoch to make the system Sybil-resistant. Epoch randomness and committee reconfiguration are some of the main components of the sharding system. In this section, we will discuss randomness generation methods followed by blockchain sharding systems and their use in committee reconfiguration. 
\subsection{Randomness Generation Methods} Generating \textit{random beacon} that satisfies the criteria of \textit{public-verifiability, unbiasness, and unpredictability} is a hard problem. There exist several protocols addressing this issue, including Verifiable Random Function \cite{vrf}, RandHound, RandHerd \cite{rand}, HydRand \cite{schindler2020hydrand}, and Verifiable Secret Sharing \cite{vss}. In this section, we will discuss these randomness generation protocols. 
\subsubsection{VRF}
In traditional pseudorandom oracles \cite{pseudo}, a secret seed is used to generate a random output for a given input, but these outputs can not be independently verified without revealing the secret seed. In this system, the owner of a secret seed $s$ can calculate the function $f_s$ at any input point $x$, resulting in a value $v$, which is the function's output for that particular input. VRF also provides NP-proof $proof_x$ which demonstrates that the output $v$ is indeed the correct result of applying the function $f_s$ to the input $x$, and it does so without giving away the secret seed $s$. This means that while the function remains verifiable for specific inputs (where the proof is provided), it still functions as a pseudorandom oracle for all other inputs, preserving the element of unpredictability. However, the proof does not require uniqueness. 
\subsubsection{RandHound} 
RandHound adopts a two-step commit-then-reveal process to generate random values. This process is facilitated by Publicly Verifiable Secret Sharing (PVSS) \cite{pvss}. To ensure the integrity of the protocol's output and prevent the client from equivocating, RandHound integrates the CoSi \cite{cosi} witnessing mechanism. In this protocol, they define a group \( G \) of large prime order \( q \) with a generator \( G \). The network comprises a set of nodes, \( N = \{0, \ldots, n-1\} \), where the nodes excluding the zeroth node, \( S = N \setminus \{0\} \), represent the servers. The system is designed to accommodate a maximum of \( f \) Byzantine nodes, necessitating that the total number of nodes, \( n \), satisfies \( n = 3f + 1 \). Each participant in the network, including the client (node 0) and servers (nodes with an index greater than 0), possesses a unique key pair. Specifically, the client has the key pair \( (x_0, X_0) \), while server \( i \) holds \( (x_i, X_i) \). The servers are organized into disjoint trustee groups, \( T_l \), each defined for \( l \) in the range \( \{0, \ldots, m - 1\} \). These groups have a secret sharing threshold \( t_l = \lceil |T_l|/3 \rceil + 1 \). The session configuration, which is publicly available, is denoted as \( C = (X, T, f, u, w) \). This includes the list of public keys \( X = (X_0, \ldots, X_{n-1}) \), the server grouping \( T = (T_0, \ldots, T_{m-1}) \), a purpose string \( u \), and a timestamp \( w \). The hash of this configuration, \( H(C) \), serves as a unique identifier for each session. It is essential that both the session configuration and the identifier are unique for every run of the protocol. All nodes are presumed to be aware of the public keys list \( X \). The output of the RandHound protocol in this setup is a random string \( Z \), which can be verified for its authenticity using a transcript \( L \). 
\subsubsection{RandHerd} RandHerd offers a continuously operating, decentralized service capable of producing randomness that is both publicly verifiable and immune to bias. This service is designed to function on-demand or at predetermined intervals. The objective of RandHerd is to minimize further the communication and computational demands associated with randomness generation, scaling down from RandHound's $O(c^2n)$ to \(O(c^2 \log n)\) for a group size $c$. This efficiency is achieved through an initial setup phase that securely divides the nodes into smaller subgroups. Utilizing aggregation and communication trees, RandHerd efficiently generates random outputs thereafter. Like before, RandHerd's random output $\hat{r}$, along with the associated challenge $\hat{c}$, is unbiased and can be authenticated as a collective Schnorr signature \cite{schnorr} against RandHerd's combined public key.
\subsubsection{HydRand}
HydRand is a protocol based on Scrape's Publicly Verifiable Secret Sharing (PVSS) \cite{scrape} designed to continuously provide random values at regular intervals, particularly in environments susceptible to Byzantine failures. The protocol ensures the generation of new randomness that is resistant to bias in every round, guaranteeing the reliability and consistency of its output. It offers a probabilistic guarantee that predicting future random values becomes exponentially difficult, enhancing the security against prediction attacks. For applications that wait for at least $f+1$ rounds before using a protocol output, HydRand provides absolute certainty in unpredictability. The protocol operates under a synchronous system model with $n=3f+1$ participants, a common setup for Byzantine fault tolerance.  Compared to previous PVSS-based approaches, HydRand reduces communication complexity from $O(n^3)$ to $O(n^2)$. 
\subsubsection{VSS}
VSS is a cryptographic method where a secret message is encrypted and divided among a set of processors, with a certain number of them (a quorum) required to access the information. The protocol supports any threshold $t$ and requires just two rounds of communication. It has low communication and computation complexity, denoted as $O(nk)$ and $O((n\log n+k)(nk\log k)$ respectively, where $k$ is a security parameter. The protocol assumes the existence of hard-to-invert encryption functions and demonstrates compatibility with discrete log encryption in finite fields, on elliptic curves, and based on r-th residues and RSA, all of which possess the required properties for the protocol.

\subsection{Committee Reconfiguration}
If the shards' nodes are static and not reshuffled after each epoch, the adversaries can control a specific shard or multiple shards to perform attacks benefiting themselves. Some adversaries may leave and join the network to be assigned to their desired shard and to take control of any specific shard or multiple shards. To prevent adaptive adversaries, the committees need to be reconfigured after each epoch, ensuring the honest majority in each shard. In this section, we will discuss the sharding protocol's committee reconfiguration protocols to make the system robust against any biasness or Sybil attack. 
\subsubsection{Elastico}
Each member of the final committee independently generates a random $r$-bit string $R_i$ and circulates the hash $H(R_i)$ among committee members. A consensus is reached on a collection of these hash values, labeled $S$, through an interactive consistency protocol. This collection, encompassing at least $\frac{2c}{3}$ of the hash values (with $c$ representing the committee's size), acts as a binding commitment to the random strings. This agreed-upon set $S$ is then shared with the entire network. Following the validation of $S$ (indicated by obtaining $\frac{2c}{3}$ signatures on it), each committee member broadcasts their respective $R_i$ to the whole network. This procedure ensures that the genuine members only disclose their commitments after confirming the committee's unanimous decision on $S$. This prevents adversaries from manipulating their commitments. Nodes in the network will receive a range of $\frac{2c}{3}$ to $\frac{3c}{2}$ pairs of $R_i$ and $H(R_i)$ from the committee, disregarding any $R_i$ that don't align with the corresponding $H(R_i)$. The finalized set $S$ is then utilized to set up the configuration for the next epoch. However, this epoch reconfiguration requires all the nodes to be reassigned to shards, which is resource-intensive due to the large overhead of bootstrapping. Besides, it is difficult to maintain individual ledgers for each shard, particularly when there is a possibility of replacing several committee members in every epoch. 
\subsubsection{OmniLedger}
To address the issue of randomness generation and committee reconfiguration, OmniLedger uses RandHound and leader selection algorithm VRF-based Algorand 
\cite{algorand}. At the start of each epoch $e$, every validator calculates a ticket using the configuration information of all the registered validators for epoch $e$. Validators elect the node with the lowest valid value as the leader for the upcoming RandHound protocol execution. If the leader fails to start the protocol, the validators slide the view window and restart the selection process. Following a successful RandHound run and the leader's distribution of $rnd_e$ along with its validity proof, each of the $n$ properly registered validators can first authenticate and then utilize $rnd_e$ to generate a permutation $\sigma_e$ of $1,\dots,n$. This permutation is then divided into $m$ roughly equal segments, determining the allocation of nodes to different shards. 
\subsubsection{RapidChain} RapidChain addresses the issue of shuffling the nodes of the shards using \textit{Cuckoo rule} \cite{cuckoo}. The reconfiguration consists of three steps: offline PoW, epoch randomness generation, and reconfiguration of the shards. The nodes who want to join the network must solve a PoW puzzle to establish their valid identity, which is done offline. The reference committee $C_r$ is responsible for validating the solutions of the new nodes and then agrees on a block, including the list of all valid nodes. RapidChain employs the VSS \cite{vss} method to produce unbiased random values within its reference committee. For node allocation to shards, the system initially assigns every node a random point within the interval [0,1) through a hashing process. Subsequently, the interval is segmented into $k$ equal shards, each measuring $\frac{k}{n}$. A committee is formed from nodes falling within $O(\log (n))$ of these shards, given a constant $k$. 
\subsubsection{TEE-based Protocol} This protocol uses \textit{epoch randomness} generated in the previous epoch to randomly re-assign the nodes to the shards. This randomness is generated using TEE. For the randomness generation, the nodes broadcast the $rnd$ generated by TEE to the network. After a certain time, the nodes select the lowest $rnd$ to be the randomness to assign the nodes to the shards. 
\subsubsection{Brokerchain} In $BrokerChain$, \textit{P-shard} performs the state graph partitioning, and after the account segmentation, a PBFT is run in the \textit{P-shard} to agree on the state block $B^t$, which contains the necessary data for the next epoch. The formation of the \textit{P-shard} and \textit{M-shards} are updated using the \textit{Cuckoo rule}. The state graph partitioning tries to reduce the number of cross-shard transactions by assigning nodes with a higher number of transactions between them in the same shard. But this poses a security risk where attackers can perform a high number of dummy transactions to be assigned in the same shard in the next epoch.

\begin{table*}[t]
    \centering
    \resizebox{\textwidth}{!}{
    \LARGE
    \begin{tabular}{||c|c|c|c|c|c|c|c|c|c||}
    \hline \hline
         &\textbf{Identity Establishment}  & \textbf{Network Model} & \textbf{Intra-Shard Consensus}   &  \textbf{Committee Configuration}    &  \textbf{Transaction Model}  & \textbf{Fault Tolerance} &\textbf{Throughput*}       &  \textbf{Latency}  \\ \hline
         \textbf{Elastico}   & PoW           & Partial Sync. & PBFT                  & Full Shuffle                  & UTXO                  & 33\%              & 16 blocks in 110s & 110s for 16 blocks \\ \hline
         \textbf{OmniLedger} & PoW           & Partial Sync. & ByzCoinX              & Rolling                       & UTXO                  & 25\%              & 13000 tps         & $\approx1s$ \\ \hline
         \textbf{RapidChain} & Offline PoW   & Partial Sync. & Adapted PBFT          & Partial Shuffle               & UTXO                  & 33\%              & 7384 tps          & 8.7 s \\ \hline
         \textbf{Monoxide}   & Public Key    & Async.        & PoW                   & Static                        & Account/Balance       & 50\%              & 11694 tps         & 16 s \\ \hline
         \textbf{TEE-based}  & TEE           & Partial Sync. & PBFT                  & Full Shuffle                  & UTXO                  & 33\%              & -                 & - \\ \hline
         \textbf{Pyramid}    & PoW           & Partial Sync. & BFT                   & Static                        & UTXO                  & 33\%              & 12000 tps         & 6 s \\ \hline
         \textbf{RepChain}   & PoW           & Partial Sync. & Raft \& BFT           & Full Shuffle                  & UTXO                  & 33\%              & 1485 tps          & 58.2 s \\ \hline
         \textbf{Brokerchain} & PoW          & -             & PBFT                  & Partial Shuffle               & Account/Balance       & 33\%              & 3000 tps          & 14.87 s \\ \hline \hline
    \end{tabular}
    }
    \caption{Comparison of sharding protocols}
    *The number of nodes in the entire network, number of shards, and shard size are unique to the experimental setup of each protocol; hence, no common ground can be found for comparing the throughput of the protocols. However, we have considered the maximum throughput obtained by each protocol under their experimental setup.
    \label{tab:comparison}
\end{table*}

\section{Performance Comparison}
Sharding protocols target various goals and configurations, often customized for specific conditions and use cases. Hence, conducting real and large-scale experiments poses the challenges of finding a common ground for meaningful performance comparison and carrying out the associated cost. Consequently, this SoK paper focuses on design rational and theoretical analysis.
In this section, we compare the performance of the state-of-the-art blockchain sharding protocols. 

Table \ref{tab:comparison} provides a comparison of the state-of-the-art public blockchain sharding protocols. \\ 
\textit{Identity Establishment} shows how new nodes can join the network. They need to solve PoW puzzles in most of the protocols, while TEE-based and only public-key-based identity establishment is also followed by some protocols. \\ 
\textit{Network Model} represents the characteristic of the underlying network. While most of the protocols follow a partially synchronous model, Monoxide follows an asynchronous model as it leverages PoW for intra-shard consensus. \\ 
\textit{Intra-shard consensus} represents the consensus mechanism the sharding protocols follow to reach an agreement upon the state of the ledger within the shard. Most of the sharding protocols follow PBFT, while OmniLedger uses ByzCoinX, which is an adopted PBFT protocol. RepChain uses Raft \cite{raft} to generate a transaction chain. \\ 
\textit{Committee configuration} represents how the nodes are assigned to the shards. \textit{static} committees do not change the assignment within the protocol. In contrast, some protocols use \textit{Cuckoo Rule} \cite{cuckoo} to shuffle the committees \textit{partially}, and other protocols fully shuffle the committee configuration after each epoch. \\
\textit{Transaction Model} shows that most of the sharding protocols use the UTXO model. \\ 
\textit{Fault Tolerance} represents the fraction of adversaries the protocol can tolerate. \\ 
\textit{Throughput} represents the number of transactions the protocols can accomplish per second. Elastico's throughput is measured in how many blocks it can generate in a certain period of time. \\ 
\textit{Latency} shows the transaction confirmation latency of the protocols. 

\section{Discussion \& Future Research Directions}
In the current sharding protocols, data migration overhead is a major limitation. After each reconfiguration of the committees, the nodes need to store the disjoint ledger of their new shards, which leads to data migration overhead. \textit{OmniLedger} addresses this issue by introducing \textit{checkpoints}. When a checkpoint is reached, the previous UTXOs are stored in a block. Therefore, the nodes are not required to retain the entire ledger from the genesis block. \textit{SSChain} \cite{sschain} introduces two-layered solution for this. In layer one, the root chain maintains security, ensuring it possesses more than half of the computational power of the whole network. The nodes can join the shards that maintain disjoint ledgers. These shards are not re-shuffled after each epoch; thus, SSChain prevents data migration. Since this protocol does not reconfigure the shards, any adversary can slowly gain control over the shards. 

Generating \textit{unbiased} and \textit{unpredictable} randomness is another major challenge for the sharding protocols to ensure the security of the shards. Leader-based random beacon generation algorithms like \textit{SPURT} \cite{das2022spurt}, \textit{OptRand} \cite{bhat2022optrand} can be explored to address this issue.

Besides the state-of-the-art sharding protocols that we have discussed, other sharding protocols like \textit{RSCoin} \cite{rscoin} employ a dual-phase commitment method. Initially, transactions are submitted to the leaders of the input shards. If these transactions gain approval from most of the input leaders, they are then forwarded to the leaders of the output shards for final verification. \textit{Chainspace} \cite{al2017chainspace} is a smart contract-based sharding protocol that uses a distributed commit protocol to guarantee the consistency of the sharding state. In Chainspace, miners within the input shards first achieve consensus among themselves. Following this, the input shard leaders engage in communications with the leaders of the output shards to establish a consensus that spans across different shards. 

The limitations in current blockchain sharding protocols and the challenges they face present significant research opportunities. Future research directions addressing these issues can be: 
\begin{itemize}
    \item \textit{Reducing Data Migration Overhead:} SSChain reduces the data migration overhead, but at the same time, it poses a security threat since it does not reconfigure the committees after each epoch. So, preventing data migration overhead while preserving the security of the shards is still an open research question. 
    \item \textit{Smart Contract-based Sharding:} Since smart contract-based blockchains can be leveraged to facilitate various functionalities, including decentralized apps (DApps) \cite{dapps}, decentralized finance (DeFi) \cite{defi}, etc., sharding in smart contract-based blockchains should be explored more to ensure better scalability. 
    \item \textit{Dynamic Shard Management:} Future studies might explore dynamic sharding mechanisms where shards can be created, merged, or dissolved based on network load and transaction patterns. This flexibility could reduce unnecessary data migration by adapting the shard structure to current network conditions.
\end{itemize}

\section{Conclusion}
In this work, we present a systemization of knowledge for public blockchain sharding. We provide an analysis of the state-of-the-art blockchain sharding protocols, focusing on their core components, including intra-shard and cross-shard transactions, consensus protocols, committee formation and reconfiguration, and identity establishment. These protocols employ various mechanisms for ensuring the security and consistency of the blockchain network, such as leader election, cross-shard communication, and data availability and integrity. Additionally, we offer insights into the key components and limitations of these protocols. Through performance comparisons and analysis of their respective consensus, fault tolerance, and identity establishment mechanisms, we provide a holistic understanding of the current landscape in public blockchain sharding. 




\printbibliography

\end{document}